# A spin-photon interface using charge-tunable quantum dots strongly coupled to a cavity

Zhouchen Luo[1], Shuo Sun[1,†], Aziz Karasahin[1], Michael K. Yakes[2], Samuel G. Carter[2], Allan S. Bracker[2], Daniel Gammon[2], Edo Waks[1*]

[1]*Department of Electrical and Computer Engineering, Institute for Research in Electronics and Applied Physics, and Joint Quantum Institute, University of Maryland, College Park, MD 20742, USA.*

[2]*Naval Research Laboratory, Washington, DC 20375, USA*

*edowaks@umd.edu

**Abstract**. Charged quantum dots containing an electron or hole spin are bright solid-state qubits suitable for quantum networks and distributed quantum computing. Incorporating such quantum dot spin into a photonic crystal cavity creates a strong spin-photon interface, in which the spin can control a photon by modulating the cavity reflection coefficient. However, previous demonstrations of such spin-photon interfaces have relied on quantum dots that are charged randomly by nearby impurities, leading to instability in the charge state, which causes poor contrast in the cavity reflectivity. Here we demonstrate a strong spin-photon interface using a quantum dot that is charged deterministically with a diode structure. By incorporating this actively charged quantum dot in a photonic crystal cavity, we achieve strong coupling between the cavity mode and the negatively charged state of the dot. Furthermore, by initializing the spin through optical pumping, we show strong spin-dependent modulation of the cavity reflectivity, corresponding to a cooperativity of 12. This spin-dependent reflectivity is important for mediating entanglement between spins using photons, as well as generating strong photon-photon interactions for applications in quantum networking and distributed quantum computing.

**Main**

Charged epitaxially-grown III-IV quantum dots that contain an electron or hole spin[1,2] have emerged as a promising solid-state qubit system. They exhibit a high radiative efficiency,[3] which is important for generating single or entangled photons of high brightness. In addition, they support fast all-optical coherent spin rotations on picosecond timescales,[4–6] necessary for high-speed quantum network and quantum information processing. Coupling such charged quantum dots to photonic crystal cavities enables strong spin-photon interfaces,[7] which are essential for solid-state quantum networks[8–10] and the generation of strong photon-photon interactions.[11–13] These applications require the spin-state of the quantum dot to modulate the cavity reflectivity, allowing the spin to control the state of the reflected photons (e.g., polarization and frequency).

Several studies have demonstrated such strong spin-photon interfaces between the electron spin of a charged quantum dot and cavity,[14–17] enabling optical nonlinearities such as Kerr rotations[16,17] and single-photon transistors.[15] These studies have relied on probabilistically charged quantum dots due to nearby impurities. However, this charging mechanism results in low charge stability due to carrier tunneling,[18] causing poor spin initialization and qubit gate fidelity.[14] Alternatively, charge tunable devices,[19–21] which typically feature a p-i-n diode structure, can be used to control and stabilize the electron charging state in a quantum dot. This diode structure can also increase atom-cavity cooperativity (a figure of merit that describes the efficacy of the coherent energy exchange between the emitter and cavity field) by suppressing spectral wandering due to electric field noise induced by trapped surface charges.[22,23] Recently, there has been significant progress in incorporating such diodes in photonic structures that exhibit small mode volumes and hence strong emitter-cavity coupling strength.[18,24–26] However, to attain a strong interface with spin remains an outstanding challenge.

Here we report a spin-photon interface using a charge-tunable quantum dot electron spin strongly coupled[27,28] to a cavity. To attain this interface, we exploit destructive interference between the cavity field and quantum dot transitions,[9,29] which are spin-state dependent, to modulate the cavity reflectivity. By optically pumping[30,31] the spin population from thermal equilibrium, we observed a significant reflectivity contrast at the cavity resonance,[14] indicating an atomic cooperativity of 12 as compared to the previously reported maximum of 2.[14,16,25] This enhanced cooperativity corresponds to a significant increase in the reflectivity contrast. Such a high-fidelity spin-photon interface could be useful for applications including quantum repeaters,[9] single photon qubit gates,[11] and deterministic spin-photon entanglement.[32]

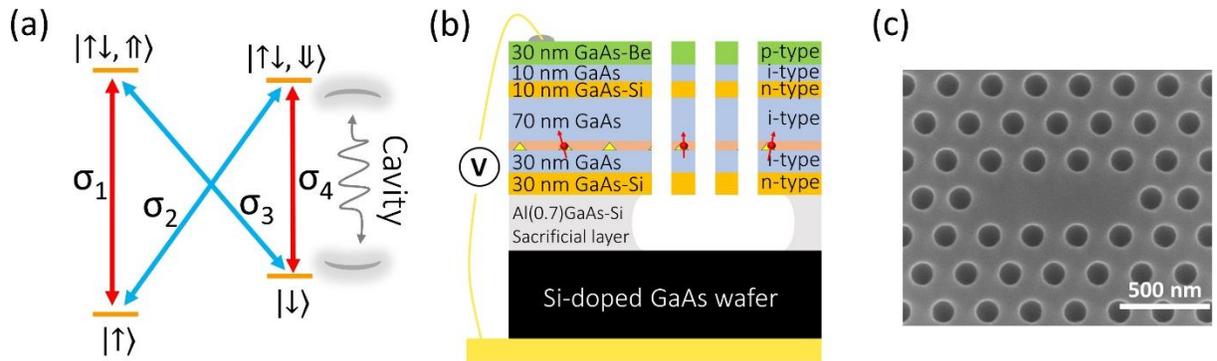

**Figure 1.** (a) The energy-level diagram of a single electron charged quantum dot coupled to a microcavity under a magnetic field in the Voigt configuration. Only transition σ4 is on-resonance with the cavity. (b) Schematic of the layered structure of the p-i-n-i-i-n GaAs diode. The yellow triangles in the center represent quantum dots and the red arrows represent electron spin. A DC bias was applied across the wafer to tune and stabilize the charging state of the quantum dots. (c) Scanning electron microscopy image of the L3 photonic crystal cavity (top view).

Figure 1(a) is the energy-level diagram of a negatively charged quantum dot under an external in-plane magnetic field (Voigt geometry). It features two opposite electron spins (spin up and spin down) that form the two ground states, labelled as $|\uparrow\rangle$ and $|\downarrow\rangle$, respectively. Meanwhile an electron pair with the opposite hole spin form the two excited states, labelled as $|\uparrow\downarrow, \Uparrow\rangle$ and $|\uparrow\downarrow, \Downarrow\rangle$.[33] An in-plane magnetic field can be applied to break the energy degeneracy of the opposite spin states, enabling us to individually address the four optically allowed and linearly polarized transitions ($\sigma_1$ to $\sigma_4$).

In order to trap and stabilize an electron in the quantum dot to form this four-level system, we utilize a device structure featuring InAs quantum dots embedded in a p-i-n-i-i-n GaAs diode membrane of 180 nm in thickness (layered structure shown in Figure 1(b)). In this membrane we patterned L3 photonic crystal cavities[34] (scanning electron microscopy image shown in Figure 1(c)) that featured a high-quality factor (~$10^4$). We used an optical cryogenic setup similar to that in Sun *et al.*[14] to measure the photoluminescence and cross-polarized reflectivity spectrum of the quantum dots and cavity. Details about the device fabrication and optical measurements can be found in the Methods.

We measured the photoluminescence spectrum of the cavity at different gate voltages to identify the charging states of the embedded quantum dots. Figure 2(a) shows the measured spectra as a function of the voltage bias. A quantum dot with a resonance frequency of ~927.7 nm (labelled $X^0$) emits at a bias voltage range of 0.64–0.74 V, while at a higher voltage range of 0.73–0.99 V we observed another emission at ~931.5 nm (labelled $X^-$). As further explained below, these states correspond to a neutral and charged exciton emission. The bright peak near 932 nm (labelled "Cavity") is the L3 cavity mode illuminated by nearby quantum dot emissions. The $X^-$ peak is much brighter than the $X^0$ peak due to Purcell enhancement by the nearby cavity mode. We focused on this particular cavity and $X^-$ state to explore its possible spin-dependent behavior.

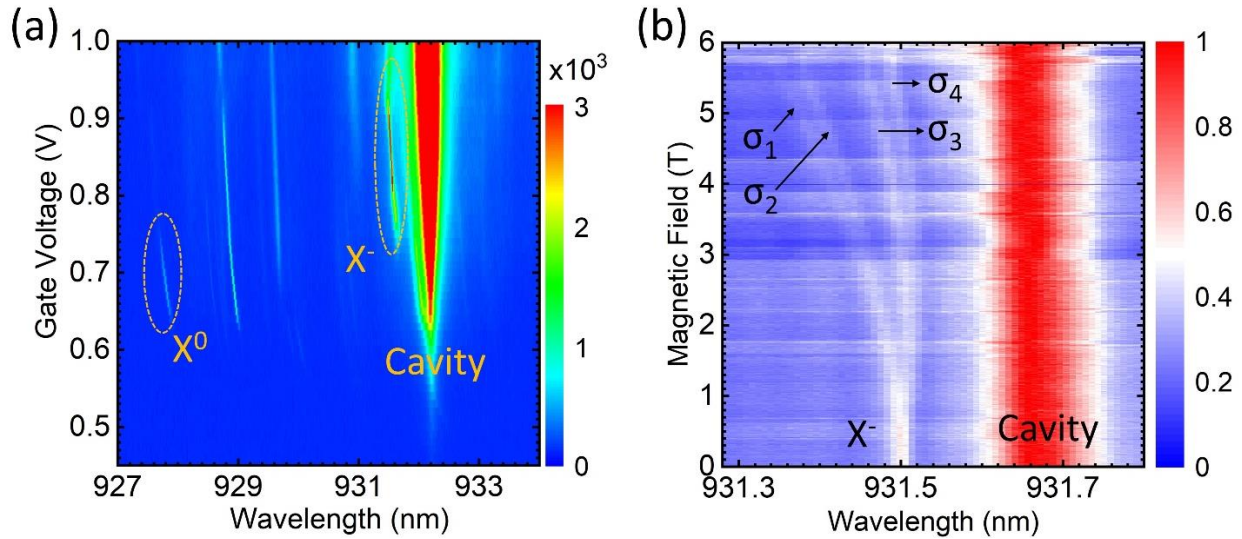

**Figure 2.** A single electron charged quantum dot near the cavity resonance. (a) Photoluminescence spectra of quantum dots and the cavity at different gate voltages. Inside the dashed circles are the neutral exciton state ($X^0$) and single electron charged state ($X^-$) of a quantum dot. (b) Cross-polarized reflectivity spectrum under different magnetic fields (0–6 T, Voigt configuration). The single emission peak of the quantum dot splits into four emission peaks with increasing magnetic field amplitude, corresponding to the four optical transitions (σ1 to σ4, from short to long wavelength) shown in Figure 1(a). We note that there are several jumps and wandering in the spectra, which are artifacts that occur because we need to periodically re-align the sample by moving the piezo stage to compensate for the sample position displacement caused by the changing magnetic field.

To determine the charging state of X⁻ in Figure 2(a), we measured its reflectivity spectrum under increasing external magnetic field (Voigt-geometry) using cross-polarized resonance reflectivity spectroscopy (Figure 2(b)).[33] For these measurements, the gate voltage was set to 0.87 V to stabilize the X⁻ state, where it features the strongest signal and narrowest linewidth (Figure 2(a)). The single X⁻ peak at low magnetic field splits into four peaks at high field (> ~3T), indicating a four-energy level system, which suggests a charged quantum dot state (i.e., a trion; Figure 2(b)).[33] We also scanned a tunable laser across this transition and observed the Raman emission from two excited states[35], which further verified the four-energy level structure (See section I of Supporting Information).

In order to determine whether the X- state is due to a hole or electron charging event, we calculated the Lande g-factor of this quantum dot using the equation $g_l = \hbar\Delta_e/\mu_B B$, in which $\Delta_e$ is the energy splitting between two trion ground states under an external magnetic strength of $B$, $\mu_B$ is the Bohr magneton. For this system, we found a Lande g-factor of $g_l \approx 0.47$, which is consistent with the 0.4–0.6 range reported for a single-electron-charged quantum dot.[4,6,14,30,36] Furthermore, the X⁰ and X⁻ states are separated by ~3.8 nm (Figure 2(a)), which is consistent with the previously reported trion binding energy of an InGaAs quantum dot.[18,24–26] This combined evidence suggests that X⁰ and X⁻ are the neutral and negatively charged states of the quantum dot, respectively.

To enable the charged quantum dot to modulate the cavity reflectivity under an external magnetic field, it was necessary to bring one of the trion transitions on-resonance with the cavity. However, trion transitions shift toward shorter wavelengths with increasing magnetic field, as shown in Figure 2(b). Therefore we shifted the cavity wavelength to the shorter side of the quantum dot emission at 0 T (~931.45 nm, Figure S2) by performing a surface cleaning treatment (see section II of the Supporting Information for more details). We then applied a magnetic field to break the trion energy degeneracy and shift one of the transitions on-resonance with the cavity *via* the Zeeman Effect to achieve coupling. Figure 3 is the reflectivity spectra as a function of this magnetic field amplitude at 0.87 V, excited by a broadband light emitting diode. In contrast to Figure 2(b), where the cavity and dot are clearly separated in emission wavelength, the emission spectrum in Figure 3 exhibits multiple anti-crossings between the different trion transitions ($\sigma_1$–$\sigma_4$) as they are tuned across the wavelength of the cavity mode due to the strong coupling between the quantum dot and the cavity.[22,27,28] At ~6 T, transition $\sigma_4$ is near on-resonance with the cavity. We note that $\sigma_3$ is also coupled to the cavity with a slight detuning under these conditions due to the small splitting between the two excited states (~0.04 nm at 6 T).

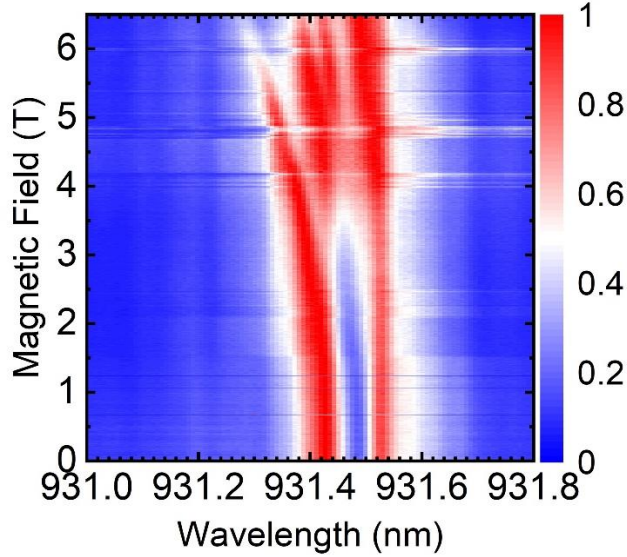

**Figure 3.** Normalized reflectivity spectra versus the Voigt-geometry magnetic field amplitude, from 0 T to 6.5 T at 0.87 V. The spectral jumps and wandering observed here are also due to sample stage realignment in response to the magnetic field, as in Figure 2(b).

We therefore fixed the magnetic field at 6.2 T where $\sigma_4$ is near on-resonance with the cavity to study how the quantum dot spin state affects the cavity reflectivity. We first measured the cavity reflectivity at thermal equilibrium (Figure 4(a)) and observed two dips in the spectrum at 931.42 nm and 931.46 nm, corresponding to transitions σ3 and σ4, respectively. These dips are due to the cavity reflectivity modification brought by the destructive interference between the quantum dot transitions and cavity field. The relative depth of those dips depends on the atomic cooperativity between the transition and the cavity, defined as $C = 2g^2/\kappa\gamma$, where $g$ is the coupling strength between the individual transition and cavity, and $\kappa$ and $\gamma$ are the decay rate of the cavity and the quantum dot transition. We measured $\kappa/2\pi = 31.79\ GHz$, corresponding to a quality factor around 10,000 (See section III of supporting information). Transitions σ1 and σ2 near 931.32 nm are too weak to be resolved in this spectrum because of large detuning from the cavity.

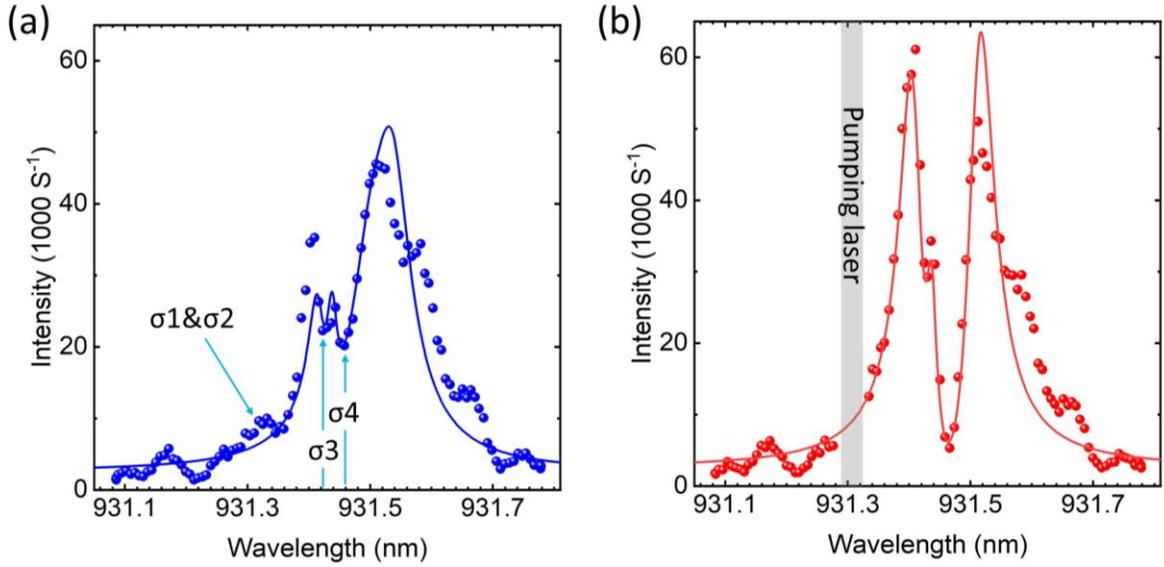

**Figure 4.** The effect of optical pumping on the spin population and cavity reflectivity at 0.87 V and 6.2 T. (a) The cross-polarized reflectivity spectrum (blue dots) measured without laser pumping. The blue solid line is the numerical fitting. The blue arrows indicate the wavelength positions of the quantum dot's four optical transitions. (b) The cross-polarized reflectivity spectrum (red dots) measured with the pumping laser on resonance with transition σ1. The data points near 931.32 nm were discarded because of the strong reflected signal from the optical pumping laser at that wavelength. The red solid line is the numerical fitting using the same model as (a).

The depths of the two dips in the reflectivity spectrum are poor because the spin state of the quantum dot is not in the pure $|\downarrow\rangle$ state but rather a mixed state of $|\uparrow\rangle$ and $|\downarrow\rangle$, with a probability of $P_{up}$ and $P_{down} = 1 - P_{up}$, respectively, determined by thermal equilibrium. Accordingly, the reflectivity spectrum of the mixed state ($R_m$) can be viewed as a probabilistic superposition of two individual spectra, $R_{up}$ and $R_{down}$, quantified as $R_m = P_{up} R_{up} + (1 - P_{up}) R_{down}$,[14] where $R_{up}$ and $R_{down}$ are the reflectivity spectra when the quantum dot spin is in the pure $|\uparrow\rangle$ or $|\downarrow\rangle$ state, respectively. $R_{up}$ can be approximated by the bare cavity reflectivity (i.e., without any quantum dot emission; Figure S3(a)), since when the spin population is in the $|\uparrow\rangle$ state, the quantum dot can only make transitions σ1 and σ2 (Figure 1(a)), which are largely detuned and thus not coupled to the cavity. When the population is in the $|\downarrow\rangle$ state, the quantum dot can only make optical transitions σ3 and σ4. Thus the reflectance at the cavity resonance wavelength will be strongly suppressed and appear as a dip in the spectrum due to the destructive interference between σ4 and the cavity field. We fit the reflectivity results in Figure 4(a) to a theoretical model (blue line) and extracted $P_{up}$ to be 0.52 (see section IV of the Supporting Information for the fitting process). While theoretically, the population occupation at thermal equilibrium should be proportional to $e^{-\Delta E/k_B T}$, from which we calculated $P_{up}$ to be 0.39, with $\Delta E \sim 0.165\ meV$ (energy splitting between $|\uparrow\rangle$ and $|\downarrow\rangle$, 0.12 nm) and $k_B T \sim 0.362\ meV$ (at 4.2 K). We attribute the deviation of the experimentally extracted value of $P_{up}$ from the theoretical prediction to the distortion of reflectivity spectrum caused by the inevitable etaloning fringes of the optical setup.

We used optical pumping[14,30,31] to move the system out of thermal equilibrium and initialize the quantum dot to the spin-down state to observe the resulting change in the cavity reflectivity. We chose to resonantly drive transition σ1 using an optimal pumping wavelength and power (931.32 nm, 120 µW; See section V of Supporting Information) because driving σ2 would introduce Raman emission near the cavity resonance wavelength, which would interfere with the reflectivity measurements. As we pumped this transition, we simultaneously probed the system's cavity reflectivity (Figure 4(b)). The dip at 931.47 nm created by the interference between the cavity and the quantum dot emission field is more significant compared to that in Figure 4(a), because the optical pumping initializes more spin population to the $|\downarrow\rangle$ state.

We fit the spectrum (red line Figure 4(b)) and extracted $P_{up} < 0.02$ (95% confidential bound), which shows we were able to initialize most of the spin population to the $|\downarrow\rangle$ state, consistent with a previous report[30]. Based on this fitting (see section IV of the Supporting Information for calculations), we also extracted the detuning between the σ4 transition and cavity to be within $2.8\ GHz$, confirming that σ4 was near resonance with the cavity. The corresponding coupling was $g_4/2\pi = 17.2 \pm 0.6 GHz$. The coupling between the σ3 transition and cavity was thus $g_3/2\pi = \sqrt{g^2 - g_4^2} = 7.2$ GHz. We believe the difference between σ3 and σ4's coupling strength with the cavity is due to the different polarization alignment of σ3 and σ4 with the cavity field polarization. From the fitting, we also extracted the dephasing of σ4 and σ3 to be $\gamma_{d4}/2\pi = 1.4 \pm 0.4\ GHz$ and $\gamma_{d3}/2\pi = 3.1 \pm 1.5 GHz$, respectively. From these derived parameters, we calculated the cooperativity between σ4 and the cavity to be $C = 2g^2/\kappa\gamma = 12.4$, which is a significant improvement from previous reports of coupled negatively charged quantum dot-cavity systems (C~2).[14,16,25] We attribute this improvement partially to the increased charging stability imposed by the diode. The system also satisfies the strong coupling criteria ($4g > \kappa + \gamma$), meaning it operates in the strong coupling regime[27,28,37]. We also tried optically pumping σ3 to transfer the population to the spin-up state. However, because the detuning between σ3 and σ4 is small (12 GHz (0.04 nm) separation), we cannot resolve the probe laser from the direct reflection of the stronger optical pumping laser.

**Conclusion**

In conclusion, we have achieved spin-dependent cavity reflectivity within a charge tunable device. The diode structure allows us to deterministically load and stabilize electron spin inside quantum dots, which leads to a cooperativity as high as 12. Using samples with a bottom distributed Bragg reflector could increase the device signal collected. Additionally, integrating the cavity with a tapered fiber[38] or on-chip waveguide[39] would allow higher signal collection efficiency. Increasing the quality factor of the L3 cavity to over 50,000[40] could also lead to a much larger atomic cooperativity. The strong solid state spin and photon interface demonstrated in this work could enable many quantum information processing tasks, such as deterministic spin-photon entanglement[32] and single-shot spin readout,[41] and could be used as a building block for future quantum networks linked by photonic qubits.

**Methods**

## Device Structure

The GaAs diode-gated InAs quantum dot material was grown by a molecular beam epitaxy process similar to that used by Carter et al.[26] We spin-coated a positive photoresist (ZEP520A) with a thickness of ~500 nm on the sample surface. L3 cavities with a lattice constant of 236 nm and hole radius of 70 nm were patterned into the photoresist by electron beam lithography and then transferred to the diode by inductively coupled plasma dry etching. A suspended cavity membrane with a thickness of 180 nm was formed after we wet-etched away the n-type AlGaAs sacrificial layer using hydrofluoric acid. A scanning electron microscopy image (top view) of one of our fabricated L3 cavities is shown in Figure 1(c). The nearest three holes to the cavity center were shifted to increase the cavity quality factor.[34] Electrical contacts were made to the top and bottom of the sample with indium and conductive epoxy, and the sample was glued to a chip carrier.

## Optical measurement setup

The sample was mounted in a closed cycle liquid helium cryostat with a base temperature of ~4.3 K. A superconducting coil allowed us to apply a magnetic field of up to 9 T at a direction perpendicular to the sample growth (Voigt configuration). For photoluminescence measurements, we excited the InAs quantum dots with an above-band laser (~890 nm) and collected their photoluminescence signal through the same objective lens with a N.A. of 0.68. For cavity reflectivity measurements, we excited the cavity with right-circularly polarized light and collected the left-circularly polarized component of the reflected signal using either a broadband light emitting diode or a narrow linewidth tunable laser for better spectral resolution. The rejection between the two cross-polarized components imposed by a pair of polarizers in the setup was larger than $10^5$. The collected signal was then directed to a spectrometer with a grating resolution of less than 0.02 nm. For the optical pumping measurement, we simultaneously sent two narrow linewidth, tunable lasers into the cavity.

## AUTHOR INFORMATION

**Corresponding Author**

edowaks@umd.edu

**Present Address**

†**Ginzton Laboratory, Stanford University, Stanford, CA 94305, USA**

**Author Contributions**

Z. L. and E. W. conceived and designed the experiments, prepared the manuscript, and carried out the theoretical analysis. Z. L. conducted the measurements and analyzed the data. S. S. contributed to the optical setup, measurements, and data analysis. A. K. contributed to sample fabrication. M. K. Y., S. G. C., A. S. B. and D. G. provided samples grown by molecular beam epitaxy.

**Funding Sources**

This work was supported by the Physics Frontier Center at the Joint Quantum Institute, the National Science Foundation (grants PHY1415485 and ECCS1508897), and the ARL Center for Distributed Quantum Information.

**Acknowledgements**

The authors declare no competing financial interests.

# Supporting Information

A spin-photon interface using charge-tunable quantum dots strongly coupled to a cavity

**Zhouchen Luo[1], Shuo Sun[1], Aziz Karasahin[1], Michael K. Yakes[2], Samuel G. Carter[2], Allan S. Bracker[2], Daniel Gammon[2], Edo Waks[1*]**

[1]*Department of Electrical and Computer Engineering, Institute for Research in Electronics and Applied Physics, and Joint Quantum Institute, University of Maryland, College Park, MD 20742, USA.*

[2]*Naval Research Laboratory, Washington, DC 20375, USA*

*edowaks@umd.edu

**Section I. Raman Emission of the Charged Quantum Dot**

To confirm the energy levels shown in Figure 1(a), we scanned the wavelength of a narrow linewidth laser through the four optical transitions at 6.2 T (Figure S1). The strong diagonal signal in Figure S1 arises from direct reflection of the scanning laser off of the sample surface. When the laser wavelength is close to σ1 and σ2, it produces Stokes emission close to σ3 and σ4 in wavelength, while exciting near the σ3 and σ4 transitions produces Anti-Stokes emission close to σ1 and σ2.[1] The wavelength difference between the scanning laser and the Raman emission it produces corresponds to the splitting between the two ground states of the quantum dot, which we determined to be 0.12 nm at 6.2 T (Figure 2(b)). From the peak wavelength of the Raman signal, we estimated that σ1 and σ2 are near 931.32 nm while σ3 and σ4 are near 931.44 nm, which is in agreement with the wavelength of these four transitions shown in Figure 2(b) of the main text. The observation of Raman emission induced by the pumping laser also confirms we can use the spin-flip process accompanying the Raman emission to initialize the spin state,[1–3] which is the so-called optical pumping technique that we used in this work.

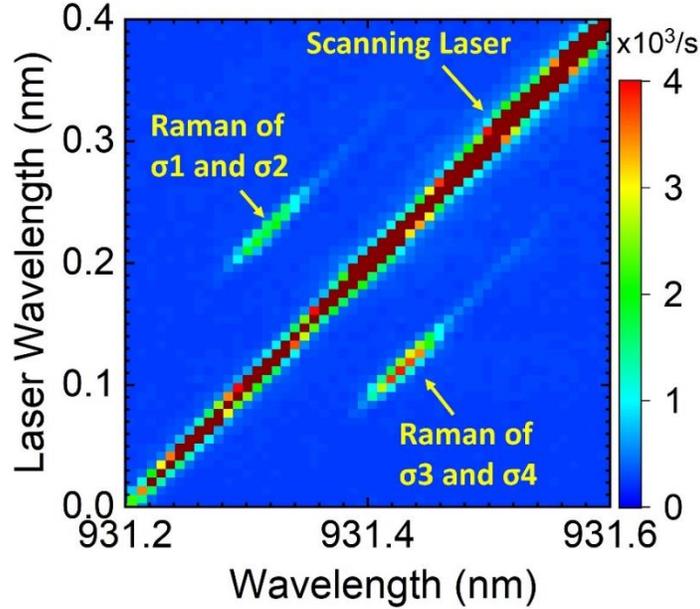

**Figure S1.** Raman emissions from the four-level charged quantum dot when scanning the pumping laser wavelength across the four optical transitions at 6.2 T and 0.87 V.

**Section II. Shifting the Cavity Wavelength**

As shown in Figure 2(b) in the main text, the trion transitions shift toward shorter wavelength under increasing external magnetic field. To bring one of the trion transitions on-resonance with the cavity when an external magnetic field is applied, we needed to blue shift the cavity resonance to the shorter wavelength side of the $X^-$ state at 0 T (931.50 nm). In order to achieve this, we flushed the sample surface with isopropyl alcohol to wash away fabrication residua, iteratively performing this cleaning step until the cavity blue-shifted to ~931.45 nm, which is near resonance but slightly blue-detuned to the $X^-$ emission.

We measured the cavity reflectivity spectra as a function of gate voltage at 0 T after we finished the cavity shifting procedure (Figure S2). The stable charging voltage range in the reflectivity spectra increased by 0.05 V compared to that in the photoluminescence spectra (Figure 2(a)). We attributed this voltage difference to the screening of the local electric field caused by free carriers created from the above-band excitation laser used for the photoluminescence measurement. Outside the stable charging voltage range, the cavity reflectivity features a Lorentzian peak (Figure S3(a)). When the quantum dot is charged (0.85–0.95 V), the cavity reflectivity spectra converts to two peaks with a dip near the cavity resonance. This strong reflection suppression near the cavity resonance is due to the destructive interference between the quantum dot dipole and cavity field,[4] as also described in the main text. As we increased the gate voltage, the wavelength of X$^-$ shifted toward shorter wavelengths due to the quantum confined Stark effect.[5]

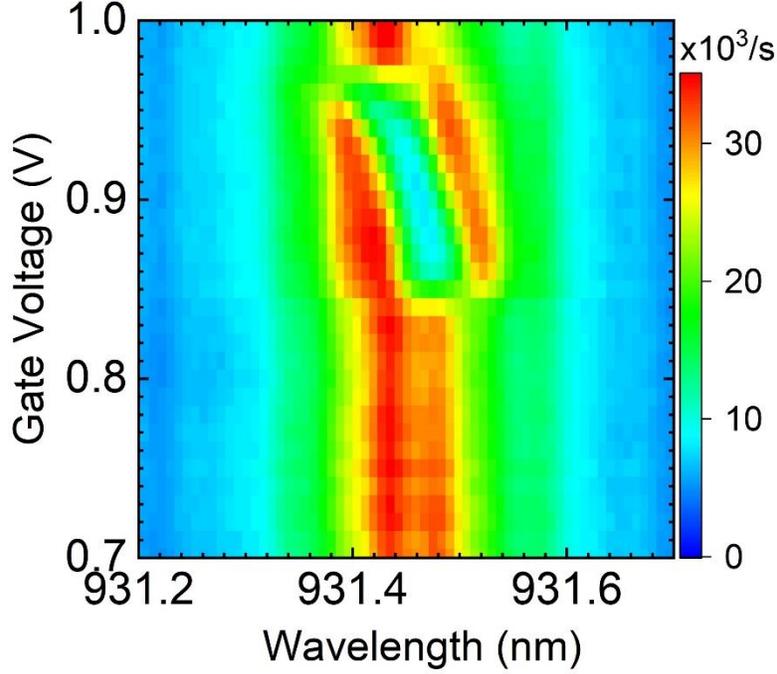

**Figure S2.** The cavity reflectivity spectra at different gate voltages excited by a broad band light emitting diode at 0 T.

**Section III. Extracting the Cavity Quantum Electrodynamic Parameters at 0 T**

To extract the cavity loss κ, we fine scanned the cavity reflectivity using a tunable narrow linewidth laser (Figure S3(a), black circles). The voltage was set to 0 V, which eliminated all the quantum dots emission (bare cavity). We fit the spectra to a Lorentzian shape (blue line in Figure S3(a)) and extracted a cavity decay rate of $\kappa/2\pi = 31.79 \pm 1.9\ GHz$, which is equivalent to a cavity quality factor of Q ~10,000.

To measure the coupling strength between the charged quantum dot and the cavity at 0 T, we scanned the cavity reflectivity at 0.87 V (Figure S3(b), black circles). We fit the coupled cavity-dot spectrum in Figure S3(b) to equation 1[4,6]:

$$R = B + S/\left|\left(-i\Delta\omega + \frac{\kappa}{2} + \frac{g^2}{(-i(\Delta\omega-\delta)+\gamma)}\right)\right|^2, \quad (1)$$

where $g$ is the coupling strength between the quantum dot and the cavity, $\gamma$ is the dipole decay rate of the quantum dot, $\Delta\omega$ is the detuning between the laser and the cavity, and $\delta$ is the detuning between the quantum dot and the cavity (red line, Figure S3(b)). $B$ and $S$ are fitting parameters accounting for the spectrum background and scaling factor, respectively. From the numerical fit we extracted g/2π = 18.67 ± 0.35 GHz and γ/2π = 1.78 ± 0.70 GHz, which gave us an atom-cavity cooperativity of $C = 2g^2/\kappa\gamma = 12.35$. The linewidth of our dot is larger than the transform limit, possibly due to the fact that the quantum dots are close to the etching surface compared with bulk quantum dots or those fabricated within micropillars.[7,8]

We note that the existence of alignment imperfection (inevitable Fabry-Perot effect among optical elements of the experiment setup) broadened the linewidth of the left peak of Figure S3(b) and the fitting tended to follow the broadened feature but couldn't fit well with the reflection dip depth. To extract an accurate cooperativity, we gave larger weight to the three data points at the center of the reflectivity spectrum of Figure S3(b) to ensure the fitting curve can fit well with the reflection dip because cooperativity is determined by the depth of the reflection dip. We also gave larger weight to the three data points at the center of reflection dip of Figure 4(b) because of the same reason.

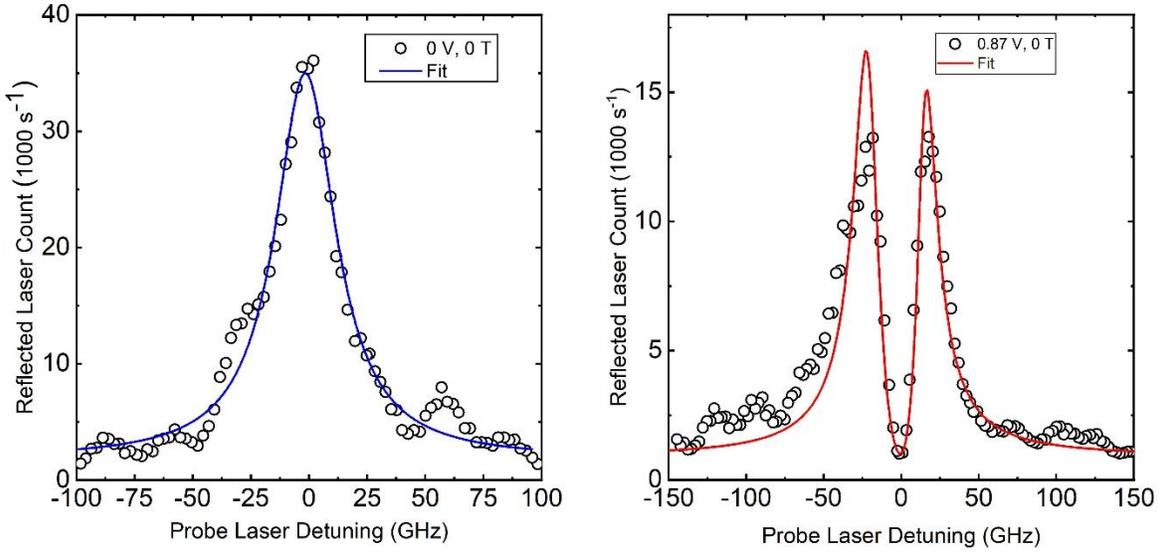

**Figure S3.** Characterization of the strongly coupled quantum dot-cavity system at 0 T. (a) The reflectivity spectrum of the bare cavity (black circles) fitted with a Lorentzian line shape (solid blue line). (b) The reflectivity spectrum of the cavity at 0.87 V. The red solid line is the theoretical fitting of the experimental data (black circles).

### Section IV. Calculating cavity quantum electrodynamics parameters and spin population distribution

To get $R_{up}(\omega)$ and $R_{down}(\omega)$, we numerically calculated the system density matrix $\rho$ at steady state governed by $d\rho/dt = -i/\hbar[\hat{H},\rho] + \hat{L}\rho$, where $\hat{H}$ is the Hamiltonian of this coupled quantum dot-cavity system when the spin is in pure $|\uparrow\rangle$ or $|\downarrow\rangle$ state and $\hat{L}$ is Liouvillian superoperator used to model all nonunitary Markovian processes in this system. $R_{up}$ and $R_{down}$ are proportional to $Tr(\rho_{ss}(\omega)\hat{a}^\dagger \hat{a})$, where $\rho_{ss}$ is the system density matrix at steady state and $\hat{a}$ is the annihilation operator of the cavity field. The spectrum of Figure 4(a) and (b) can be fit using $R_m = S * (P_{up} * Tr(\rho_{ss,\uparrow}(\omega)\hat{a}^\dagger \hat{a}) + (1 - P_{up}) * Tr(\rho_{ss,\downarrow}(\omega)\hat{a}^\dagger \hat{a})) + B$. S and B are the fitting parameters account for scaling and background as in Section III.

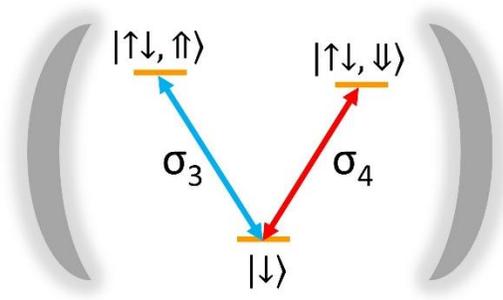

**Figure S4**. The energy-level diagram of a three-level V-scheme system coupled to a cavity, simplified from the four-energy level scheme of a charged quantum dot, shown in Figure 1(a).

When the spin population is in spin-down state, the energy level structure of the charged quantum dot can by simplified as a three level V-scheme system coupled to a cavity, as shown in Figure S4. The Hamiltonian of this system in the rotation frame with respect to the probe laser frequency can be written as:

$$\hat{H} = \hbar(\omega_c - \omega)\hat{a}^\dagger\hat{a} + \hbar(\omega_x - \omega)\hat{\sigma}_3^\dagger\hat{\sigma}_3 + \hbar(\omega_x - \Delta_h - \omega)\hat{\sigma}_4^\dagger\hat{\sigma}_4$$
$$+ ig_3\hbar(\hat{a}\hat{\sigma}_3^\dagger - \hat{\sigma}_3\hat{a}^\dagger) + g_4\hbar(\hat{a}\hat{\sigma}_4^\dagger + \hat{\sigma}_4\hat{a}^\dagger),$$

(1)

in which $\hat{\sigma}_3$ and $\hat{\sigma}_4$ are the lowering operators of transitions σ3 and σ4, respectively, $\hat{a}$ is the annihilation operator of the cavity field, $g_3$ and $g_4$ are the coupling strength between σ3 and σ4 with the cavity, respectively, $\omega_c$, $\omega_x$, $(\omega_x - \Delta_h)$ and $\omega$ are the frequency of the cavity, transition σ3, transition σ4, and the laser, respectively, and $\Delta_h$ is the Zeeman splitting between the two excited states.

The Liouvillian superoperator $\hat{L}$, which accounts for the decay of the cavity field, spontaneous emission, and dephasing of the excited trion states, can be written as:

$$\hat{L} = \kappa D(\hat{a}) + \gamma_3 D(\hat{\sigma}_3) + \gamma_4 D(\hat{\sigma}_4) + 2\gamma_{d3} D(\hat{\sigma}_3^\dagger\hat{\sigma}_3) + 2\gamma_{d4} D(\hat{\sigma}_4^\dagger\hat{\sigma}_4),$$

(2)

in which $D(\hat{O})\rho = \hat{O}\rho\hat{O}^\dagger - 1/2\hat{O}^\dagger\hat{O}\rho - 1/2\rho\hat{O}^\dagger\hat{O}$ is the general Linblad form operator for an operator $\hat{O}$, the coefficients $\kappa, \gamma_3, \gamma_4, \gamma_{d3}$, and $\gamma_{d4}$ are the cavity loss, spontaneous emission rates of σ3 and σ4, and the pure dephasing rates of σ3 and σ4, respectively.

When the spin population is in spin-up state, the quantum dot can only couple to cavity through σ2 and σ1, which are largely detuned. In this case, the system can be simplified as a bare cavity without any coupling to the quantum dot, which is equivalent to setting $g_3$ and $g_4$ to be 0 in equation (1).

We first fit Figure 4(b) using the above numerical model. We already determined $\kappa/2\pi$ to be 31.79 GHz from Figure S3(a). $\gamma_3/2\pi$ and $\gamma_4/2\pi$ were set to 0.1 GHz.[9] We also imposed $g_3^2 + g_4^2 = g^2$, in which $g$ is the coupling strength between the quantum dot and cavity at 0 T and 0.87 V, which was extracted in Figure S3(b). This constraint relation comes from the fact that the dipole transition between the excited state and ground state at 0 T ($\sigma_\pm$) could be written as $\sigma_\pm = \hat{\sigma}_4 \pm i\hat{\sigma}_3$. When multiple dipole transitions are coupled to a cavity, the overall effective coupling strength $g$ is related to individual coupling strength $g_i$ by $g^2 = \Sigma g_i^2$.[10] We used the rest system parameters $S$, $B$, $\omega_c$, $\omega_x$, $g_3$, $\gamma_{d3}$, and $\gamma_{d4}$ as free fitting parameters. We got $g_3/2\pi = 7.2\ GHz$, $g_4/2\pi = 17.2\ GHz$, $\gamma_{d3}/2\pi = 3.1\ GHz$, $\gamma_{d4}/2\pi = 1.4\ GHz$. $P_{up}$ was determined to be less than 0.02 by 95% confidential bound. The fitting result is further discussed in the main text.

With the value of $g_3$, $g_4$, $\gamma_{d3}$, and $\gamma_{d4}$ fixed from fitting of Figure 4(b), we then fit the spectrum of Figure 4(a). The spin population at thermal equilibrium was determined to be $P_{up} = 0.52 \pm 0.04$.

## Section V. Determining the optimal optical pumping conditions

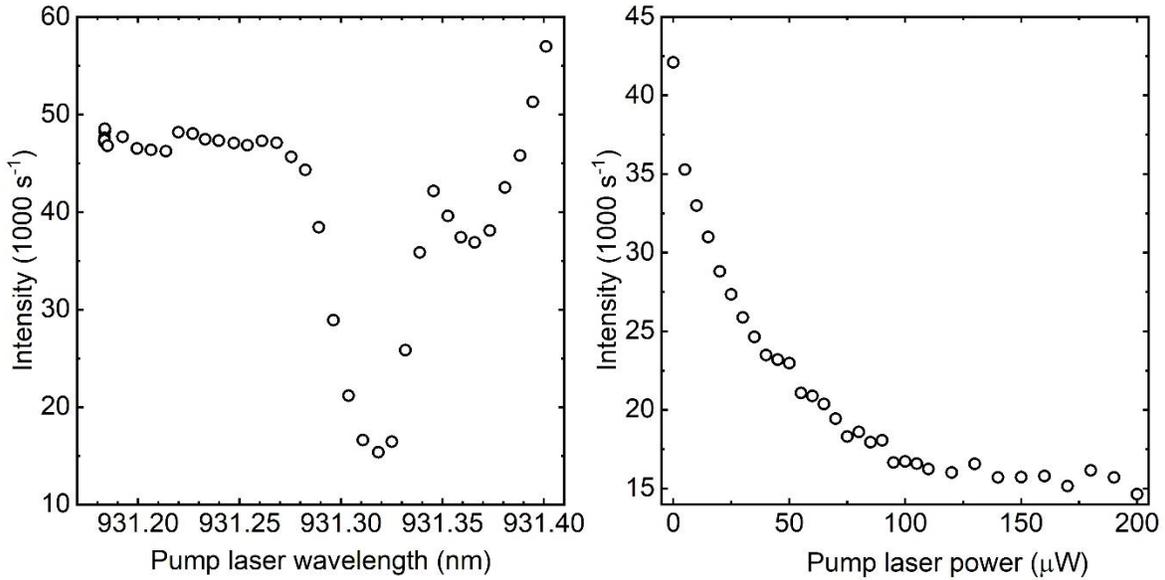

**Figure S5**. Determining the optimal optical pumping conditions. (a) The reflected probe laser (931.46 nm) count as we swept the pumping laser wavelength though transition σ1. (b) The reflected probe laser (931.46 nm) count under different pumping laser powers.

To determine the exact wavelength of σ1 at 0.87 V and 6.2 T, which we were unable to accurately determine from Figure 3 or S1, we scanned the optical pumping laser wavelength around 931.3 nm and monitored the reflected intensity of the probe laser, whose wavelength was fixed at the cavity resonance (931.46 nm; Figure S5(a)). The reflected probe laser count reached its minimum when the pumping laser was at 931.32 nm and a local minimum when the pumping laser was at 931.36 nm. We therefore concluded that the σ1 and σ2 transitions occur at 931.32 nm and 931.36 nm, respectively. As we increased the pumping laser wavelength further, the Raman emission from the pumping laser overlapped with the probe laser and the reflected signal at the wavelength of the probe laser increased sharply. This result also shows that the increase in cavity reflectivity contrast is a resonant effect that only occurs when pumping transitions σ1 and σ2, which reinforces that the increase in the dip contrast is due to an optical pumping effect.

To determine the minimum optical pumping power to initialize the spin state, we scanned the power of the optical pumping laser (fixed at 931.32 nm) and monitored the reflected probe laser intensity (Figure S5(b)). As the optical pumping power increased, it pumped a larger fraction of the spin population to the spin-down state, resulting in a deeper reflection dip and lower probe reflected intensity. The reflected probe intensity ceased to decrease after around 120 μW, which corresponds to the saturation of the spin population in the spin-down state. We thus determined 120 μW to be the minimum pumping power to initialize the quantum dot to the spin-down state.